\documentclass[a4,12pt,epsf]{article}

\begin{document}
\begin{center}{\Large \bf
Scherk-Schwarz twist in 5D conformal SUGRA
}
\end{center}

\begin{center}
Hiroyuki Abe~\footnote{abe@yukawa.kyoto-u.ac.jp}
and
Yutaka Sakamura~\footnote{sakamura@het.phys.sci.osaka-u.ac.jp}
\vspace{6pt}\\

{\it 
$^1$Yukawa institute for theoretical physics, Kyoto University, 
Kyoto 606-8502, Japan \\
$^2$Department of Physics, Osaka University, 
Toyonaka, Osaka 560-0043, Japan 
}
\end{center}
\begin{abstract}
We reinterpret the Scherk-Schwarz (SS) boundary condition for 
$SU(2)_R$ in a compactified five-dimensional (5D) 
Poincar\'e supergravity in terms of the twisted $SU(2)_U$ 
gauge fixing in 5D conformal supergravity. 
In such translation, only the compensator hypermultiplet is 
relevant to the SS twist, and various properties of the SS 
mechanism can be easily understood. 
Especially we show the equivalence between the SS twist 
and boundary constant superpotentials 
at the full supergravity level. 
\end{abstract}

\section{Introduction}
The Scherk-Schwarz (SS) mechanism~\cite{Scherk:1978ta} of SUSY 
breaking has been revisited as a phenomenologically 
interesting candidate for the physics beyond the standard 
model~\cite{Marti:2001iw}. The simplest setup in such context was 
constructed within the framework of five-dimensional (5D) supergravity 
compactified on an orbifold $S^1/Z_2$. 
Here we reinterpret the SS boundary 
condition for $SU(2)_R$ in the compactified 5D 
Poincar\'e supergravity as the twisted 
$SU(2)_U$ gauge fixing in the 5D conformal 
supergravity~\cite{Abe:2005}.\footnote{ 
This article is based on talks by Y.~Sakamura at SUSY 2006 (UC Irvine, U.S.A) 
and Summer Institute 2006 (POSTECH, Korea). }
In such an interpretation, only the 
compensator hypermultiplet is relevant to the SS twist. 

\section{$\bf SU(2)_U$ gauge fixing and Scherk-Schwarz twist}
\subsection{Twisted $\bf SU(2)_U$ gauge fixing}
In the derivation of 5D Poincar\'e supergravity from the 
5D conformal supergravity using the hypermultiplet 
compensator~\cite{Kugo:2000af,Fujita:2001bd,Kugo:2002js}, 
the $SU(2)_R$ symmetry is defined as the diagonal subgroup of 
the direct product of the original 
$SU(2)_U$ gauge symmetry among the superconformal symmetries 
and $SU(2)_C$ which rotates the compensator hyperscalars 
$({\cal A}^1_{\;\;i},{\cal A}^2_{\;\;i})$ 
($i=1,2$ is an $SU(2)_U$-index), 
\begin{eqnarray}
SU(2)_U \times SU(2)_C \to SU(2)_R, 
\label{eq:su2usu2r}
\end{eqnarray}
through the $SU(2)_U$-gauge fixing 
${\cal A}^a_{\;\; i} \propto \delta^a_{\;\;i}$. 
The dilatation- and $SU(2)_U$-gauge 
fixings completely fix 
the quaternionic compensator hyperscalars as
\begin{eqnarray}
{\cal A}^a_{\;\; i} \equiv \delta^a_{\;\; i} 
\sqrt{1+{\cal A}^{\underline\alpha}_{\;\; i} 
{\cal A}_{\underline\alpha}^{\;\; i}}. 
\label{eq:norufix}
\end{eqnarray}

However, if we consider a torus compactification of the fifth 
dimension with the radius $R$, we have inequivalent 
classes of the $SU(2)_U$ gauge fixing 
which are parameterized by a twist vector 
$\vec\omega=(\omega_1,\omega_2,\omega_3)$ as 
\begin{eqnarray}
{\cal A}^a_{\;\; i} \equiv 
\left( e^{i\vec\omega \cdot \vec\sigma\,f(y)} \right)^a_{\;\; i} 
\sqrt{1+{\cal A}^{\underline\alpha}_{\;\; j} 
{\cal A}_{\underline\alpha}^{\;\; j}}, 
\label{eq:twtufix}
\end{eqnarray}
where $y$ is the coordinate of the fifth dimension and 
$f(y)$ is a function satisfying $f(y+2\pi R)=f(y)+2\pi$. 
Nonzero $\omega_{1,2}$ correspond to the SS twist parameter.\footnote{
The consistency with the orbifold projection requires that $\omega_3=0$. }
In other words, 
from Eq.~(\ref{eq:su2usu2r}), the SS boundary condition 
for all the fields with $SU(2)_R$ index in the Poincar\'e 
supergravity is simply (equivalently) given by 
the twisted gauge fixing condition for $SU(2)_U$ in the framework 
of the conformal supergravity~\cite{Abe:2004ar}. 

Next we derive SUSY breaking terms induced by the SS twist 
in the periodic basis. By the following field redefinition, 
the compensator fixing condition~(\ref{eq:twtufix}) reduces to 
the normal one~(\ref{eq:norufix}).  
\begin{eqnarray}
{\cal A}^a_{\;\; i} \to U^a_{\ b}(y) {\cal A}^b_{\;\; i}, \qquad 
\zeta^a \to U^a_{\ b}(y) \zeta^b, 
\label{eq:bb}
\end{eqnarray}
where $\zeta^a$ is the compensator hyperinos and 
\begin{eqnarray}
U^a_{\ b}(y) \equiv 
\left( e^{-i\vec\omega \cdot \vec\sigma\,f(y)} \right)^a_{\ b}. 
\nonumber
\end{eqnarray}
As compensation for it, 
we have additional $\vec\omega$ dependent terms which arise 
from the $y$-derivatives of the compensator fields in the action. 
They are given by 
\begin{eqnarray}
e^{-1}{\cal L}_\omega
&=& 
f'(y) (i\vec\omega \cdot \vec\sigma)_{ij} \bigg\{ 
2i\bar\zeta^{\underline\beta} \gamma^4 \zeta^{\underline\alpha} 
{\cal A}_{\underline\beta}^{\ j} 
{\cal A}_{\underline\alpha}^{\ i} 
(1+{\cal A}_{\underline\alpha}^{\ k}
{\cal A}^{\underline\alpha}_{\ k})^{-1} 
+4i\bar\psi_m^i \gamma^4 \gamma^m 
\zeta^{\underline\alpha} {\cal A}_{\underline\alpha}^{\ j} 
\nonumber \\ &&
+\Big(2i\bar\psi_m^{(i} \gamma^{m4n}\psi_n^{j)}
-2{\cal A}^{\underline\alpha (i} \nabla_4 
{\cal A}_{\underline\alpha}^{\ j)} 
+ia_{IJ} \bar\Omega^{Ii} \gamma_4 \Omega^{Jj} \Big) 
(1+{\cal A}_{\underline\alpha}^{\ k}
{\cal A}^{\underline\alpha}_{\ k}) \bigg\} 
\nonumber \\ &&
-2(f'(y)|\vec\omega|)^2 
\big( {\cal A}_{\underline\alpha}^{\ i}
{\cal A}^{\underline\alpha}_{\ i}
+({\cal A}_{\underline\alpha}^{\ i}
{\cal A}^{\underline\alpha}_{\ i})^2 \big), 
\nonumber
\end{eqnarray}
after integrating out the auxiliary fields. 
This contains the mass terms of the gravitinos~$\psi^i_m$, 
the gauginos~$\Omega^{Ii}$ 
and the physical hyperscalars~${\cal A}^{\underline{\alpha}}_{\;\; i}$. 

\subsection{Singular gauge fixing and boundary interpretation}
\label{sec:sscs}
An explicit function form of $f(y)$ in Eq.(\ref{eq:twtufix}) 
does not affect the physical consequence because $f(y)$ is 
just a gauge fixing parameter. 
We usually choose $f(y)=y/R$ which gives the 
simplest description. 
However in this section, motivated by the argument of 
{\it generalized symmetry breaking} in Ref.~\cite{Bagger:2001qi}, 
we choose it as 
\begin{eqnarray}
f(y) &=& \frac{\pi}{2} \sum_n 
\big( {\rm sgn}(y-n \pi R) - {\rm sgn}(-n \pi R) \big), 
\label{eq:singularufix} 
\end{eqnarray}
where ${\rm sgn}(y)$ is the sign-function. 
Namely, 
\begin{eqnarray*}
f'(y) &=& \pi \sum_n \delta(y-n \pi R). 
\end{eqnarray*}
For this $f(y)$, ${\cal L}_\omega$ totally becomes boundary terms. 
In fact, we can show that the SS twist reproduces the same action as 
the untwisted case with the constant superpotentials~$W$ at 
both the orbifold boundaries, provided that 
\begin{equation}
 W=\pi (\omega_2+i\omega_1). 
\end{equation} 
%
From this correspondence between the SS twist and the constant 
superpotentials, we confirm that SUSY breaking caused by the SS 
twist is not explicit because 
the boundary constant superpotential is $N=1$ invariant. 
We remark that this correspondence has been easily found 
at the full supergravity level, not in the effective theory, 
thanks to our simplified interpretation of the SS twist.

\subsection{Scherk-Schwarz twist and AdS$_5$ geometry}
\label{sec:ssads5}
It was suggested in Ref.~\cite{Hall:2003yc} 
that the SS twist yields an inconsistency 
in the supergravity on AdS$_5$ geometry. 
In this section, we see how this fact can be seen 
within our framework of the twisted $SU(2)_U$ fixing. 
It is known that the gauging of $U(1)_R$ symmetry by the graviphoton 
is necessary to realize the AdS$_5$ geometry keeping 
SUSY~\cite{Altendorfer:2000rr,
Gherghetta:2000qt,Bergshoeff:2000zn}. 
In fact, the negative cosmological constant is proportional 
to the $U(1)_R$ gauge coupling in such a case. 
Thus the covariant derivatives of the compensator hyperscalars are
\begin{eqnarray}
{\cal D}_\mu {\cal A}^a_{\ i} &=& 
\partial_\mu {\cal A}^a_{\ i} 
-(g_R W_\mu^R)^a_{\ b} {\cal A}^b_{\ i}+\cdots, 
\label{eq:covdercomp}
\end{eqnarray}
where 
\begin{eqnarray}
(g_R W_\mu^R)^a_{\ b} &\equiv& 
g_R W_\mu^R  (\vec{q} \cdot i\vec\sigma)^a_{\ b} 
\ = \ \left\{ 
\begin{array}{ll}
|\vec{q}|g_R W_\mu^R (i\sigma_1 \sin \theta_R 
+i\sigma_2 \cos \theta_R)^a_{\ b}
& (g_R:\textrm{$Z_2$-even}) \\
|\vec{q}|g_R W_\mu^R (i\sigma_3)^a_{\ b} & (g_R:\textrm{$Z_2$-odd}) 
\end{array} \right..
\nonumber
\end{eqnarray}
Since ${\cal A}^a_{\;\; i}$ are fixed to constants 
at the leading order for the gravitational coupling~\footnote{
We have taken the unit of $M_5=1$, where $M_5$ is the 5D Planck mass. } 
(see Eq.(\ref{eq:norufix})), the nonvanishing mass term for 
the graviphoton comes out from the kinetic terms 
for ${\cal A}^a_{\;\; i}$ if the commutator 
$[\vec{q}\cdot\vec{\sigma},\vec{\omega}\cdot\vec{\sigma}]$ does not vanish. 
Such a mass term breaks the unitarity of the theory because 
the graviphoton is the gauge field in this case. 
Therefore the following condition must be satisfied 
for the theory to be consistent. 
\begin{eqnarray}
\left[ \vec{q} \cdot \vec\sigma,\, 
\vec{\omega} \cdot \vec\sigma \right] &=& 0. 
\label{eq:ccfrg}
\end{eqnarray}

To obtain the GP-FLP~\cite{Gherghetta:2000qt} 
(BKVP~\cite{Bergshoeff:2000zn}) model for a supersymmetric 
warped brane world, we need to gauge the $U(1)_R$ 
symmetry by the graviphoton with $Z_2$-odd gauge coupling\footnote{
The $Z_2$-odd gauge coupling can be realized in the supergravity 
through the four-form mechanism~\cite{Bergshoeff:2000zn}.}
$g_R$, i.e. $\vec{q}=|\vec{q}|(0,0,1)$.  
From the condition~(\ref{eq:ccfrg}), 
the only possible twist vector in this case is $\vec\omega=0$. 
Namely the SS twist is impossible in this case. 

On the other hand, in the ABN model~\cite{Altendorfer:2000rr} 
in which the $U(1)_R$ symmetry is gauged by the graviphoton with 
$Z_2$-even gauge coupling $g_R$, i.e. 
$\vec{q}=|\vec{q}|\,(\sin \theta_R,\cos \theta_R,0)$, 
the possible twist vector is 
$\vec\omega=|\vec\omega|\,(\sin \theta_R,\cos \theta_R,0)$. 
So we find a possibility to have the SS twist 
in the warped spacetime. 
However it was pointed out 
that ABN model is not derived from the known off-shell formulations 
with the linear multiplet~\cite{Zucker:2000ks} 
or the hypermultiplet compensator~\cite{Kugo:2002js}. 
We can not find the Killing spinor on this background 
in those off-shell formulations for ABN model 
even without the SS twist. 
This is still an open question. 

\section{Conclusion}
By noticing that the twisted $SU(2)_R$ boundary condition 
in $S^1/Z_2$-compactified 5D Poincar\'e supergravity is equivalent 
to the twisted $SU(2)_U$ 
gauge fixing in 5D conformal supergravity, we have reexamined 
the SS twist boundary condition in the latter terminology. 
In this case, only the compensator hypermultiplet is relevant 
to the SS twist, and various properties of the SS mechanism 
can be easily understood. 
We reproduced the 5D Poincar\'e supergravity with the SS twist 
from the 5D conformal 
supergravity with the twisted $SU(2)_U$ gauge fixing. 
Thanks to this interpretation, we can explicitly show 
the Wilson line interpretation of the SS twist~\cite{vonGersdorff:2001ak}, 
the correspondence between the SS twist and 
the boundary constant superpotentials~\cite{Bagger:2001qi}, 
and the quantum inconsistency of the 
twist in the AdS$_5$ background~\cite{Hall:2003yc} 
{\it at the full supergravity level}. 

\pagebreak

\vspace*{12pt}
\noindent
{\bf Acknowledgement}
\vspace*{6pt}

\noindent
The authors would like to thank the organizers of 
SUSY 2006 and Summer Institute 2006. 



\begin{thebibliography}{9}
\bibitem{Scherk:1978ta}
  J.~Scherk and J.~H.~Schwarz,
  Phys.\ Lett.\ B {\bf 82}, 60 (1979).

\bibitem{Marti:2001iw}
  D.~Marti and A.~Pomarol,
  Phys.\ Rev.\ D {\bf 64}, 105025 (2001)
  [hep-th/0106256]; 
%
  D.~E.~Kaplan and N.~Weiner,
  hep-ph/0108001.

\bibitem{Abe:2005}
 H.~Abe and Y.~Sakamura, JHEP~{\bf 0602}, 014 (2006) [hep-th/0512326].

\bibitem{Kugo:2000af}
  T.~Kugo and K.~Ohashi,
  Prog.\ Theor.\ Phys.\  {\bf 105}, 323 (2001)
  [hep-ph/0010288].

\bibitem{Fujita:2001bd}
  T.~Fujita, T.~Kugo and K.~Ohashi,
  Prog.\ Theor.\ Phys.\  {\bf 106}, 671 (2001)
  [hep-th/0106051].

\bibitem{Kugo:2002js}
  T.~Kugo and K.~Ohashi,
  Prog.\ Theor.\ Phys.\  {\bf 108}, 203 (2002)
  [hep-th/0203276].

\bibitem{Abe:2004ar}
H.~Abe and Y.~Sakamura,
JHEP {\bf 0410}, 013 (2004)
[hep-th/0408224].

%

\bibitem{Bagger:2001qi}
  J.~A.~Bagger, F.~Feruglio and F.~Zwirner,
  Phys.\ Rev.\ Lett.\  {\bf 88}, 101601 (2002)
  [hep-th/0107128]; 
%
  J.~Bagger, F.~Feruglio and F.~Zwirner,
  JHEP {\bf 0202}, 010 (2002)
  [hep-th/0108010]; 
%
  C.~Biggio, F.~Feruglio, A.~Wulzer and F.~Zwirner,
  JHEP {\bf 0211}, 013 (2002)
  [hep-th/0209046].

\bibitem{Hall:2003yc}
  L.~J.~Hall, Y.~Nomura, T.~Okui and S.~J.~Oliver,
  Nucl.\ Phys.\ B {\bf 677}, 87 (2004)
  [hep-th/0302192].

\bibitem{Altendorfer:2000rr}
R.~Altendorfer, J.~Bagger and D.~Nemeschansky,
Phys.\ Rev.\ D {\bf 63}, 125025 (2001)
[hep-th/0003117].

\bibitem{Gherghetta:2000qt}
T.~Gherghetta and A.~Pomarol,
Nucl.\ Phys.\ B {\bf 586}, 141 (2000)
[hep-ph/0003129]; 
%
A.~Falkowski, Z.~Lalak and S.~Pokorski,
Phys.\ Lett.\ B {\bf 491}, 172 (2000)
[hep-th/0004093].

\bibitem{Bergshoeff:2000zn}
E.~Bergshoeff, R.~Kallosh and A.~Van Proeyen,
JHEP {\bf 0010}, 033 (2000)
[hep-th/0007044].

\bibitem{Zucker:2000ks}
M.~Zucker,
Phys.\ Rev.\ D {\bf 64}, 024024 (2001)
[hep-th/0009083]; 
%
M.~Zucker,
Fortsch.\ Phys.\  {\bf 51}, 899 (2003).

\bibitem{vonGersdorff:2001ak}
  G.~von Gersdorff and M.~Quiros,
  Phys.\ Rev.\ D {\bf 65}, 064016 (2002)
  [hep-th/0110132]; 
%
  G.~von Gersdorff, M.~Quiros and A.~Riotto,
  Nucl.\ Phys.\ B {\bf 634}, 90 (2002)
  [hep-th/0204041]; 

\end{thebibliography}
\end{document}